\documentclass[journal]{IEEEtran}
\usepackage{graphicx}
\usepackage{color}
\usepackage{cite}

\if11
\usepackage[a4paper,breaklinks=true]{hyperref} 
\usepackage{breakurl}
\hypersetup{
     bookmarks=true,         
     unicode=false,          
     pdftoolbar=true,        
     pdfmenubar=true,        
     pdffitwindow=false,     
     pdfstartview={FitH},    
     pdftitle={Demonstrator of the Belle II Online Tracking and Pixel Data Reduction on the High Level Trigger System},    
     pdfauthor={T. Bilka, G. Casarosa, R. Fruhwirth, C. Kleinwort, P. Kodys, P. Kvasnicka, J. Lettenbichler, E. Paoloni, J. Rauch, T. Schluter, S. Yashchenko},     
     pdfsubject={Belle II High Level Trigger Data Reduction},   
     pdfnewwindow=true,      
     colorlinks=false,       
     linkcolor=red,          
     citecolor=green,        
     filecolor=magenta,      
     urlcolor=cyan           
}
\fi

\newcommand{\belle}{Belle~II}
\newcommand{\genfit}{\textsc{Genfit}}

\begin{document}

\title{Demonstrator of the \belle{} Online Tracking and Pixel Data Reduction on the High Level Trigger System}
%
%

\author{T.~Bilka, G.~Casarosa, R.~Fr\"uhwirth, C.~Kleinwort, P.~Kody\v{s},
  P.~Kvasni\v{c}ka, J.~Lettenbichler, E.~Paoloni, J.~Rauch, T.~Schl\"uter,
  S.~Yashchenko%
  \thanks{Manuscript received June 16, 2014; revised December 22, 2014.}
  \thanks{This research was supported by the DFG cluster of excellence
    'Origin and Structure of the Universe', by the Ministry of
    Education, Youth and Sports of the Czech Republic under Contract
    No. LG14034, by the Austrian Science Fund (FWF), project
    P24182-N16 and under BMBF Contract 05H12WM8.  The research
    leading to these results has received funding from the European
    Commission under the FP7 Research Infrastructures project AIDA,
    grant agreement no.~262025.}
  \thanks{T. Bilka, P. Kvasni\v{c}ka and
    P. Kody\v{s} are with Faculty of Mathematics and Physics, Charles
    University in Prague, Czech Republic}%
  \thanks{R. Fr\"uhwirth and J. Lettenbichler are with HEPHY Vienna,
    Austria.
  }%
  \thanks{C. Kleinwort and S. Yashchenko are with DESY, Hamburg,
    Germany}%
  \thanks{E. Paoloni and G. Casarosa are with INFN and the University
    of Pisa, Italy}%
  \thanks{J. Rauch is with Technische Universit\"at M\"unchen,
    Germany}%
  \thanks{T. Schl\"uter as corresponding author is with Excellence
    Cluster Universe, Lud\-wig-Maxi\-mi\-lians-Uni\-ver\-si\-t\"at
    M\"unchen, Garching, Germany (email:
    tobias.schlueter@physik.uni-muenchen.de).}%
}
\maketitle
\thispagestyle{empty}

\begin{abstract}
  The future \belle{} experiment will employ a computer-farm based
  data reduction system for the readout of its innermost detector, a
  DEPFET-technology based silicon detector with pixel readout.  A
  large fraction of the background hits can be rejected by defining a
  set of Regions Of Interest (ROI) on the pixel detector sensors and
  then recording just the data from the pixels inside the ROI.  The
  ROIs are defined on an event by event basis by extrapolating back
  onto the PXD the charged tracks detected in the outer trackers (a 4
  layer double-sided silicon strip detector surrounded by a wire
  chamber).  The tracks are reconstructed in real time on the High
  Level Trigger (HLT).  The pixel detector is then read out based on
  the ROI information.  A demonstrator of this architecture was under
  beam test earlier this year in DESY (Hamburg, Germany).  The
  demonstrator was operated in an electron beam whose momentum was in
  the 2\,-\,6\,GeV/\textit{c} range with a typical trigger rate of a
  few kHz in a magnetic field of strength up to 1\,T.  The
  demonstrator consists of one pixel sensor and 4 silicon strip
  sensors arranged in a 5 layers configuration mimicking the \belle{}
  vertex detector.  The detector readout was a scaled down version of
  the full \belle{} DAQ + HLT chain.  The demonstrator was used to
  detect the particles, reconstruct in real time the trajectories,
  identify the ROIs on the PXD plane and record the PXD data within.
  We describe the requirements and the architecture of the final
  system together with the results obtained with the demonstrator.
\end{abstract}


\section{Introduction}
%
%
%
%
\IEEEPARstart{T}{he} \belle{} experiment is a $B$-factory experiment
currently being set up at the site of the High Energy Accelerator
Research Organization (KEK) in Tsukuba, Japan, and will take its first
physics data in 2016.  It will exploit the unprecedented luminosity
($8\times 10^{35}\,\textrm{cm}^{-2}\textrm{s}^{-1}$) of the SuperKEKB
accelerator, a fifty-fold increase over the predecessor KEKB.  The
\belle{} experiment will collect a data set of exclusive $B\bar B$
events that is two orders of magnitude larger than that of the
predecessor experiments {\sc BaBar} and
Belle~\cite{Abe:2010gxa,Bevan:2014iga}.

For many existing studies from $B$-factories, larger data sets are
desirable.  Collecting the huge amount of data
required for precision physics poses a number of challenges that must
be overcome in the experiment.  First, the accelerator needs to
reliably provide high instantaneous luminosity.  The concurrent
increase in physics data leads to a similar increase in the amount of
data read out, subject to specific choices of triggers.  The increase
in luminosity is achieved on the one hand by roughly doubling the beam
currents, on the other hand by a novel nano-beam scheme which leads to
a beam size of order $10\,\mu\textrm{m} \times 60\,\textrm{nm}$ at the
interaction point.  The increased beam current and the more delicate
optics lead to increased background levels in the detectors which are
situated close to the beam line, the second challenge.  These have to
be dealt with at the high trigger rates foreseen.  Implemenation of
the nano-beam scheme requires a reduced center-of-mass boost compared
to the previous $B$ factory experiments~\cite{Abe:2010gxa}.  In order
to compensate the thus reduced spatial separation of the $B$-decay
vertices, the vertex detector is placed very close to the interaction
point, again leading to increased background levels.

\begin{figure}[!t]
  \centering
  \includegraphics[width=\linewidth]{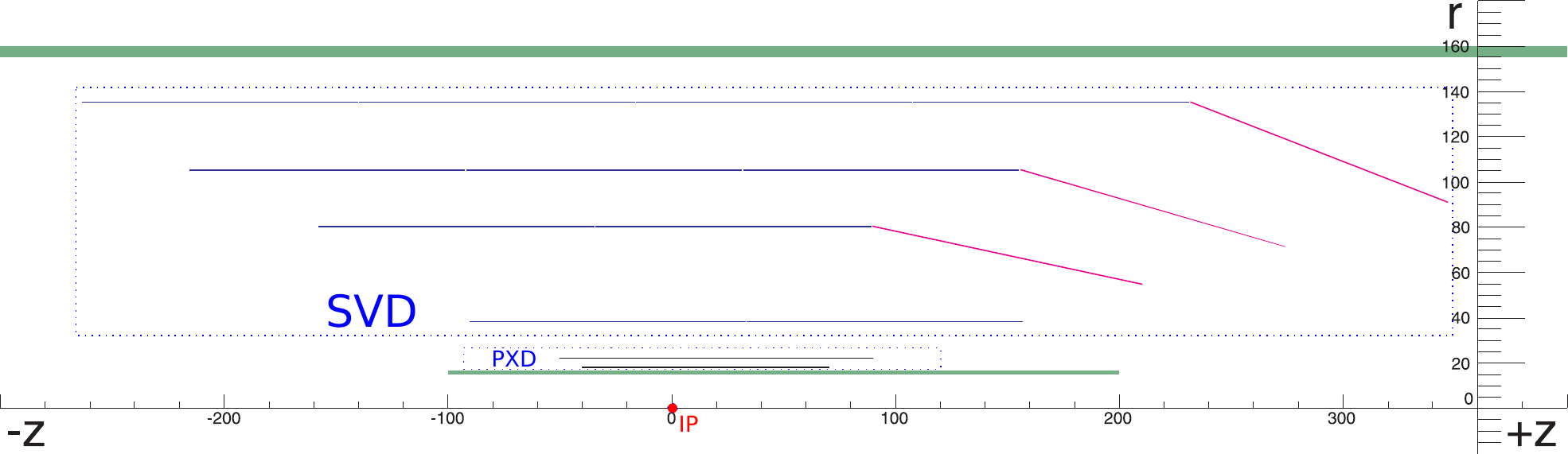}
  \caption{Layout of the \belle{} vertex detector (VXD).  Distances in
    mm relative to the interaction point (IP) are indicated.  The two
    layers of the pixel detector (PXD) and the four layers of the
    silicon vertex detector (SVD) are indicated.  The VXD volume is
    bounded by the beam pipe on the inside and by the inner wall of
    the drift chamber on the outside (both indicated in green).}
  \label{fig:vxd-layout}
\end{figure}
\begin{figure}[!t]
  \centering
  \includegraphics[width=\linewidth]{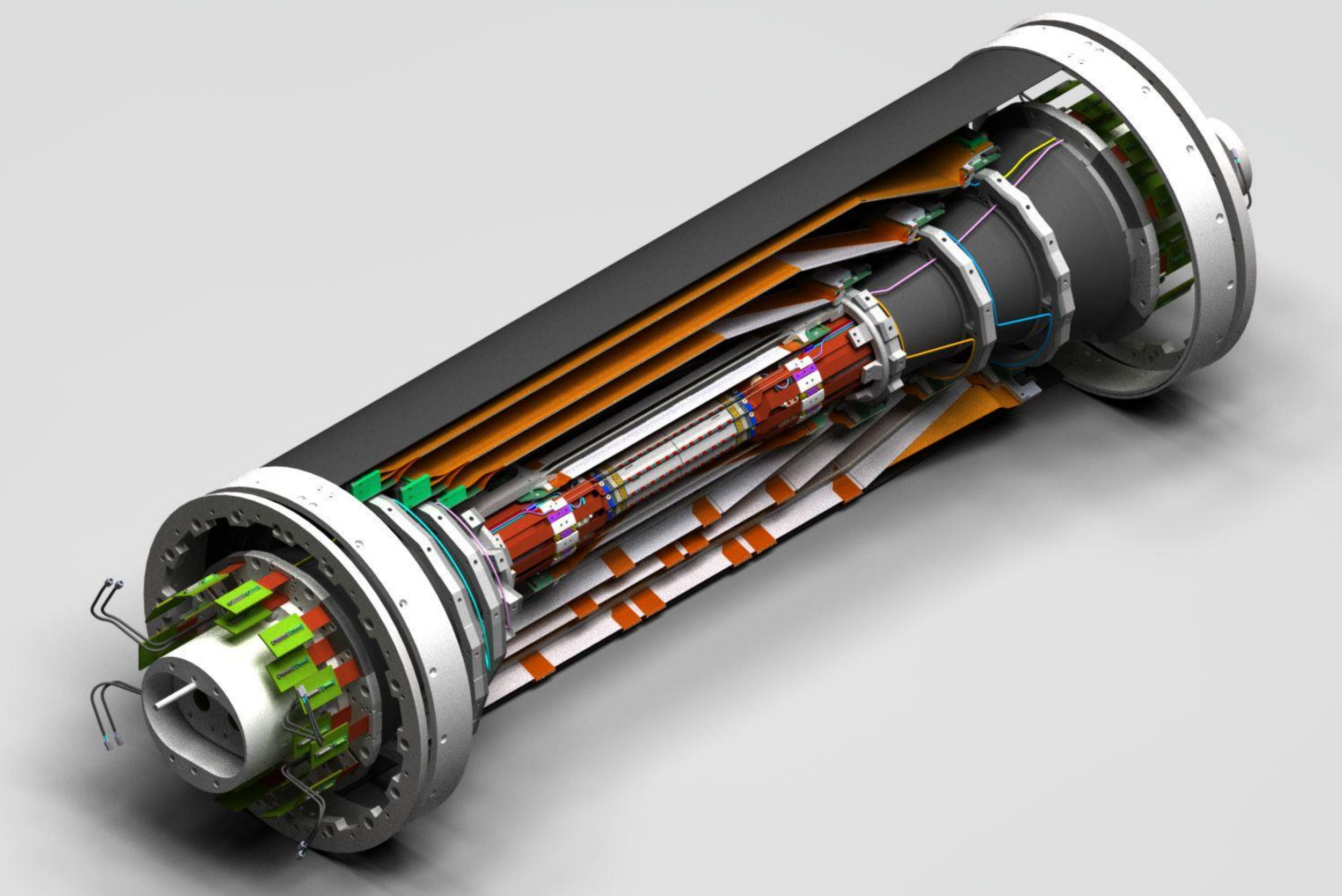}
  \caption{Illustration of the full vertex detector setup.
    Reader-facing parts of the detector have been removed for
    visibility.}
  \label{fig:vxd-layout-3d}
\end{figure}

The Vertex Detector (VXD) is composed of six layers of semiconductor
detectors arranged on concentrical cylinders around the beam pipe.
The two innermost layers (PXD) of the VXD consist of DEPFET-type pixel
sensors~\cite{Alonso:2012ss}, situated at radii of $14\,\textrm{mm}$
and $22\,\textrm{mm}$ from the beam line, respectively.  The outer
four layers (SVD) consist of double-sided silicon strip
detectors~\cite{Friedl:2013gta}, situated at radii ranging from
$38\,\textrm{mm}$ to $140\,\textrm{mm}$ from the beam line.  In order
to cover the full polar angular range of the remainder of the
experiment, $17^\circ < \vartheta < 150^\circ$, while significantly
reducing the necessary amount of silicon, the three outermost layers
employ a wedged geometry, where the part of the detector in the boost
direction is angled towards the beam line.  The VXD is surrounded by a
large drift chamber (CDC) used for precise momentum determination of
charged particle tracks and particle identification by energy loss
measurements.  Further detectors for particle identification and
neutral particle detection surround this ensemble of tracking
detectors~\cite{Abe:2010gxa}.  The layout of the VXD is shown in
Figs.~\ref{fig:vxd-layout} and~\ref{fig:vxd-layout-3d}.

Due to the proximity of the VXD, and especially the PXD, to the beam
line, sophisticated background rejection has to be employed.  Given
the expected average occupancy of the PXD of 3\%, the amount of data
would be overwhelming, exceeding 20 Gb/s of raw pixel data, or roughly
ten times the amount of data recorded by the remainder of the
experiment.  Background suppression therefore has to take place before
storing data to disk.  Additionally, the PXD is read out in frames
with an integration time of $20\,\mu s$.  This is only a factor 5
below the average spacing of two triggered events at the expected
trigger rate of $10\,\textrm{kHz}$~\cite{Abe:2010gxa}.  Therefore, a
single read-out frame can contain hits from more than one triggered
event.  These have to be correctly associated to the respective
events.

In order to achieve these objectives, the High-Level Trigger (HLT) of
\belle{} performs online track finding and fitting in the SVD and the
CDC on standard Linux PCs using the same software as in the offline
processing.  The HLT extrapolates the resulting tracks back to the
PXD, and defines Regions-of-Interest (ROIs) based on the intersections
of these tracks with the PXD surfaces.  Data contained in these ROIs
is then read out and stored for offline processing.  In parallel, the
data-merging hardware performs track-reconstruction and ROI definition
in an FPGA-based approach, which is developed as a fallback in case
implementation of the HLT approach should not work out.  For the final
experiment, this approach will reduce the amount of recorded PXD data
twenty-fold.

In this contribution we discuss the design and implementation of the
software used in the online reconstruction of the HLT system and the
results we obtained in a test beam experiment which took place in
January of 2014.  A sector of the VXD was exposed to an electron beam
at DESY, and for the first time the whole system of PXD and SVD was
successfully integrated with the \belle{} DAQ system.  The processing
chain including the above-mentioned online reconstruction could be
established during the test beam experiment.  Additionally, the
detector was operated in a cooled environment, and slow control was
handled with the slow control system foreseen for the final
experiment.

A number of other contributions at this conference dealt with other
parts of the \belle{} DAQ. The global picture of the \belle{} DAQ
system is given in S.~Yamada's contribution OS~1-1; the ONSEN system
is discussed in more detail in T.~Ge\ss{}ler's contribution
PS~3-1~\cite{Gessler:2014gba}; the event-building process is explained
in S.Y.~Suzuki's contribution PS~3-56; the design and implementation
of the digitization and readout hardware of the PXD is detailed in
D.~Levit's contribution PS~3-20~\cite{Levit:2014bba}; R.~Itoh's
contribution PFS~2 discusses more of the data processing.

\section{Experimental Setup}

\begin{figure}[!t]
  \centering
  \includegraphics[width=\linewidth]{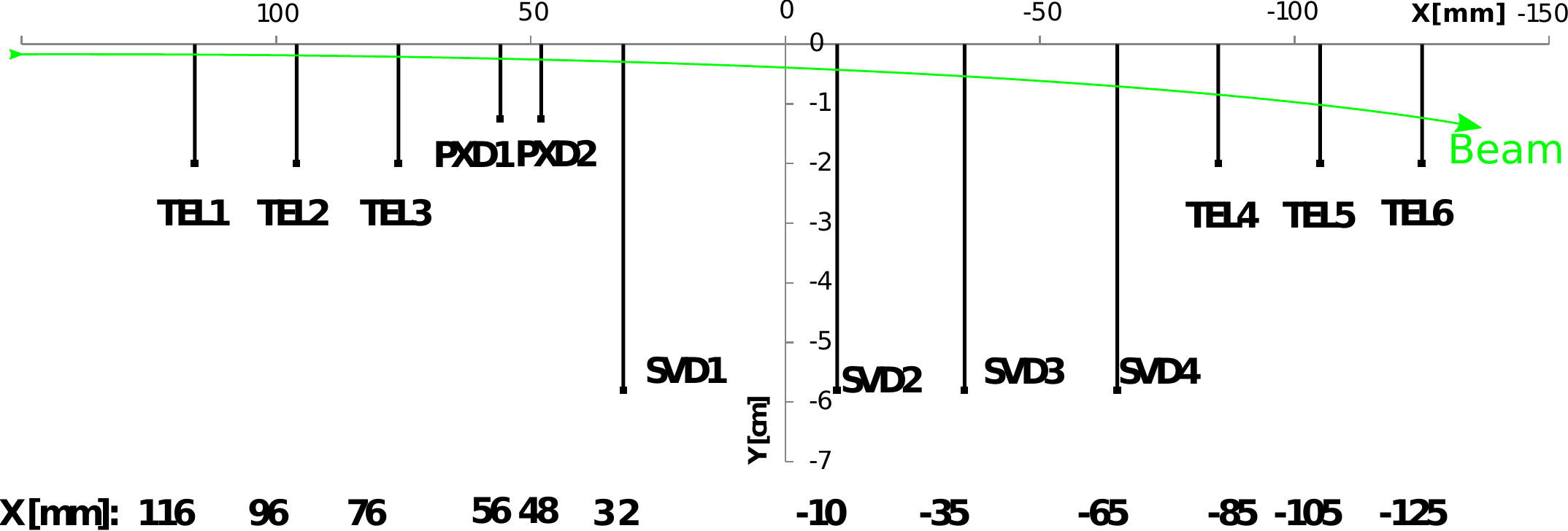}
  \caption{Experimental setup in the test beam experiment.  The beam
    (green) enters from the left where it transverses the coil of the
    solenoid magnet before reaching the detector assembly.  There it
    passes through three planes of the EUDET beam
    telescope~\cite{Bulgheroni:2010zz}.  These are followed by the VXD
    planes in a setup mimicing a sector of the final \belle{}
    assembly: first two layers of DEPFET pixel detectors (PXD1-2),
    then four planes of double-sided silicon strip detectors (SVD1-4).
    Finally, three additional telescope layers allow precise measurement
    of high-momentum trajectories.  The PXD1 plane was not present.}
  \label{fig:setup}
\end{figure}

The demonstrator was operated in an electron beam at DESY.  The
average beam momentum was in the $2 - 6\,\textrm{GeV}/ c$ range with a
typical beam rate of a few kHz.  The beam passed through the coil of a
solenoid magnet (1\,T) inside of which the detector assembly was
installed.  Bremsstrahlung processes in the magnet coil lead to a
broad momentum distribution of the particles entering the detector
volume.  A side view illustrating the detector positions and
dimensions is shown in Fig.~\ref{fig:setup}.

Unlike the full VXD setup, only one PXD layer (corresponding to PXD2
in the figure) was installed.  It contains a matrix of $480\times 192$
pixels of $75\times 50\,\mu\textrm{m}^2$ each and is thinned down to
$50\,\mu\textrm{m}$ in the active area.  As in the final experiment,
the long side is parallel to the magnetic field, the bending direction
is along the short side.  Digitization and readout ASICs are as close
as possible to the final specifications for the \belle{} experiment,
but in the final experiment the detectors will be $75\,\mu\textrm{m}$
thick.  A second sensor of the final specifications could not be
prepared in time.  Nevertheless, this setup allowed exercising the
complete readout and processing chain with real data.

Downstream of the PXD layers, four SVD layers were installed.  The
first layer had an active area of $123\times 38\,\textrm{mm}^2$ and a
thickness of $320\,\mu\textrm{m}$.  The detector had $768\times 512$
strips, with pitches of $160\,\mu\textrm{m}$ and $75\,\mu\textrm{m}$
respectively.  It was oriented in the same manner as the PXD layer.
The other three SVD layers had the same specifications, but $768\times
768$ strips and a corresponding active area of $123\times
58\,\textrm{mm}^2$.  The SVD layers employed were previously subjected
to tests of radiation hardness.  This lead to localized hot areas
which were masked during clusterization.

\begin{figure}[!t]
  \centering
  \includegraphics[width=\linewidth]{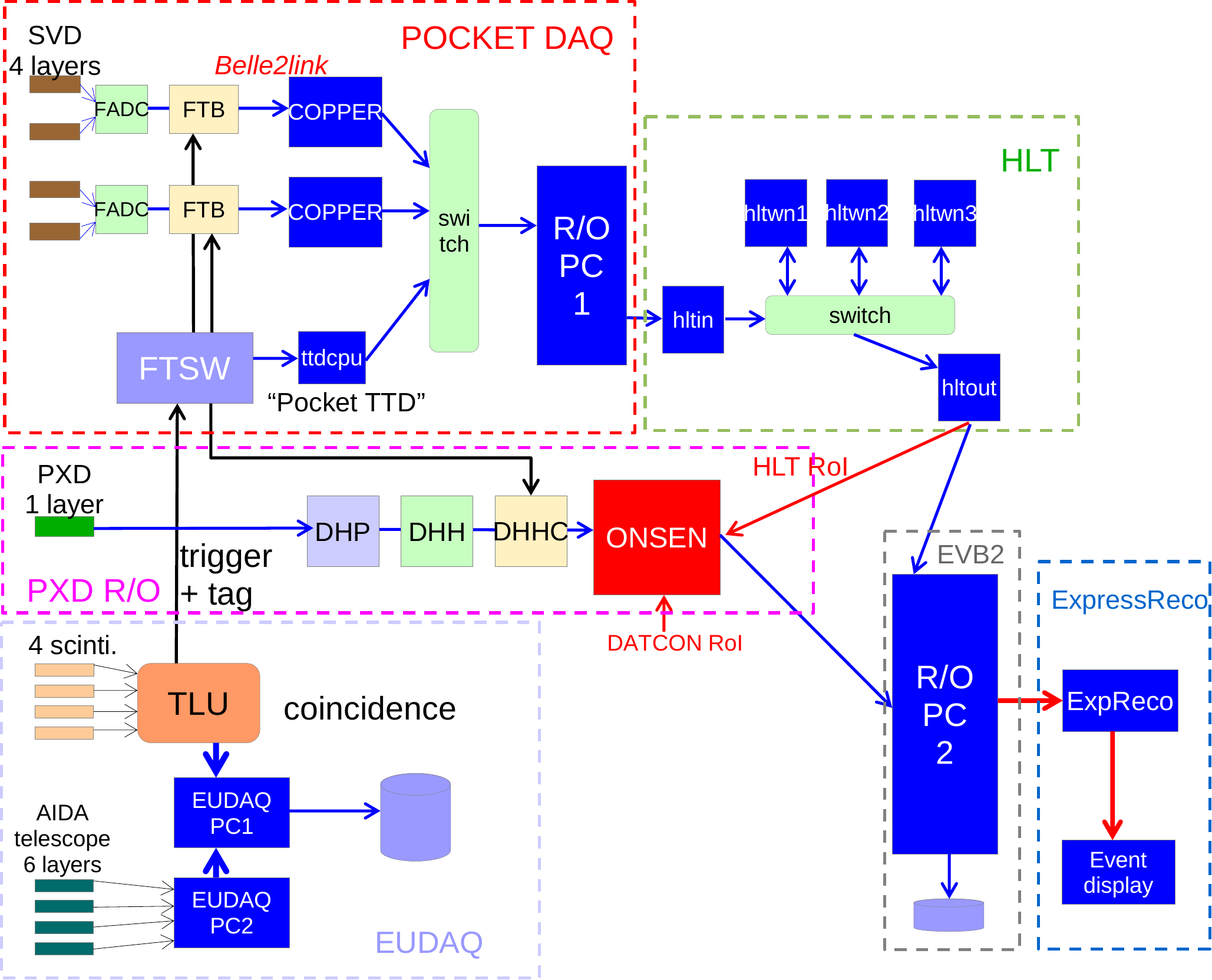}
  \caption{Block diagram of the readout of the DESY test beam
    experiment.  The software described in this contribution resides
    in the boxes labelled HLT and ExpressReco.  SVD data collected by R/O
    PC 1 is forwarded to the HLT.  Here, several PCs process data with
    event-wise parallelism.  The resultant data are forwarded to the
    second-stage event building EVB2.  At the same time the HLT sends
    ROIs to the ONSEN system which in turn forwards the pixel data
    contained in these ROIs to EVB2.  There the events are merged and
    stored to disk.  For monitoring purposes, events are also sent to
    the ExpressReco system, which performs a reconstruction including
    all layers.  (Figure courtesy of R.~Itoh.)}
  \label{fig:DAQ}
\end{figure}

The VXD assembly was contained inside a dry volume cooled with CO$_2$.
Cooling as well as detector slow control was using the final \belle{}
software.  The cold volume is placed between the two arms of the EUDET
beam telescope~\cite{Bulgheroni:2010zz} provided by the test beam
facility.  It was used for alignment purposes and efficiency studies.
The system was triggered by a coincidence between two pairs of
scintillators placed in front and behind the assembly, respectively.
The telescope data was recorded along a separate data path and merged
offline.  The readout architecture is displayed in Fig.~\ref{fig:DAQ}.
We highlight a few details in the following.

Once the Trigger Logic Unit (TLU) finds a coincidence between the
scintillators, it assigns an event number and sends it together with a
trigger signal to the telescope readout (bottom left) and the Frontend
Timing Switch (FTSW).  The FTSW is part of the Pocket DAQ (top left),
a scaled down version of the common DAQ used by all \belle{} detectors
save the PXD.  Once the Pocket DAQ has collected the data and an event
is built, it sends the data to the High Level Trigger (HLT, top right)
whose operation will be discussed in detail below.  The PXD readout
(middle left) is also initiated by the FTSW.  Upon receiving a trigger
from the FTSW, the DHHC ASIC forwards data integrated by the PXD to
the Online Selector Node (ONSEN), a buffer which has sufficient memory
to store several seconds of data under typical conditions in the
\belle{} experiment.  The ONSEN system buffers the PXD data until the
HLT (discussed below) or an FPGA-based tracking implemented in the SVD
readout hardware (not depicted) finish calculating Regions of Interest
(ROIs), i.e. parts of the detector where tracks are expected to
intersect the PXD.  Once the ONSEN system receives the list of ROIs
for an event, it forwards the corresponding pixel data to the second
event builder (bottom right) which merges the PXD data with the events
received from the HLT.  The data is then stored to disk.
Additionally, a fraction of events is forwarded to the ExpressReco, a
fast reconstruction with the purpose of data quality inspection.  This
operation matches the data paths foreseen for the \belle{} experiment.

\section{Requirements}

In the test beam experiment, the following requirements had to be
met by the HLT software:
\begin{enumerate}
\item reading of detector data formats;
\item efficient track finding and fitting;
\item extrapolation of tracks to PXD planes;
\item definition of ROIs;
\item successful communication of ROIs to the ONSEN systems.
\end{enumerate}
It was thus a critical part of the test of the complete data-handling
scheme.

The HLT system consisted of five computing nodes in total.  Two
twelve-core systems (hltin and hltout) handled the event-wise data
distribution to the HLT worker nodes (hltwn1-3) which were 24-core
systems.

The data distribution and processing was implemented in the \belle{}
software framework, {\tt basf2}~\cite{Itoh:2012hb,Lee:2011za}.  This
framework has a modular architecture, where individual data processing
steps are implemented in a chain of independent modules.  The means of
communication between the modules is the so-called datastore, which
consists of named and typed arrays where modules can read and write
data.  Subsequent processing steps can read the datastore arrays
provided by previous steps.  The implementation is such that the
datastore can be streamed to and from disk or over the network at any
point in the processing chain.  This way, processing tasks can be
distributed over a network of computers or multiple processor cores
with no further requirements on the implementation of individual
steps.  In the HLT in particular these facilities are used to
distribute the processing of the individual events over the HLT
network while concentrating communication tasks on separated nodes
(hltin and hltout in Fig.~\ref{fig:DAQ}).  Common configuration data
such as the detector geometry is held in a database, which was
implemented by XML files during the test beam.

The software employed on the HLT is the same software as is used for
offline reconstruction.  Indeed, all software used during the testbeam
were the then-current versions of the software for the \belle{}
experiment.

\section{HLT Data Processing Chain}

\subsection{Overview}

\begin{figure}[!t]
  \centering
  \includegraphics[width=.7\linewidth]{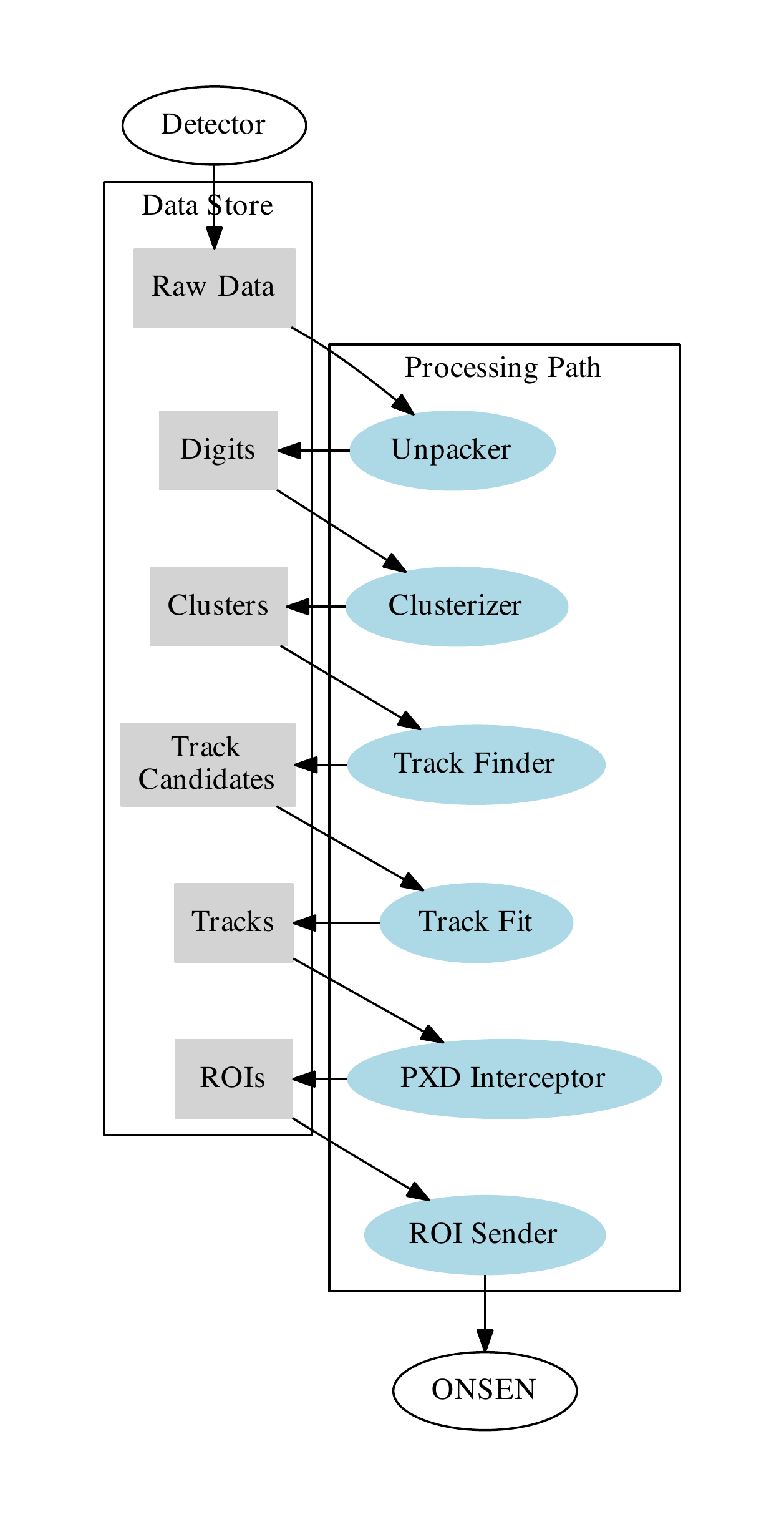}
  \caption{Block diagram of the data processing in the HLT.  The
    sequence of modules in the processing path is traversed in order.
    Each processing module is indicated as a blue oval.  Data store
    arrays (indicated as grey rectangles) serve as input and output of the
    individual modules.  Arrows indicate the direction of the data flow.}
  \label{fig:dataflow}
\end{figure}

The HLT implementation in the test beam consisted of the following
processing chain on each worker node (Fig.~\ref{fig:dataflow}): an
unpacker for the SVD data format converts the raw data to digits.  It
takes care of the conversion of channel numbers to row and column
numbers and it also masks hot areas of the detectors.  In the next
step, the digits are processed by clusterization algorithms.  These
combine the digits into clusters, evaluating the actual hit position
while taking into account Lorentz corrections to electron drift in the
presence of a magnetic field.  The clusters form the actual input for
the track finder, discussed below.  The track finder outputs track
candidates which in turn are processed with a Kalman fitter (discussed
below).  Finally the PXD Interceptor module uses the fitted tracks to
define the ROIs on the PXD.  At this point the processed events is
forwarded to the hltout node.  It runs the ROI Sender module which
takes care of the communication of the HLT with the ONSEN system.

\subsection{Track Finding}

Tracks are found in the SVD by a cellular automaton
(CA,~\cite{FruHwirth:2013jta}). The combinatorial problem of
clustering hits to track candidates is mitigated by sequentially applied filters
with increasing complexity. The cells of the CA are track segments that connect two
hits in adjacent layers. In order to reduce the number of cells, only compatible 
hits are combined to cells; this is the first filter stage. The compatibility of
a pair of hits is determined by a look-up table, called the sector map, which is
created by simulating a large number of tracks in the SVD. Two sector maps were used
in the test beam experiment, one with and one without magnetic field.

Cells sharing a hit are defined to be neighbors if they fulfill certain
geometrical requirements, for instance a cut on the angle between the
cells; this is the second filter stage. The appropriate cut values are again
obtained from the sector map. In the CA, each cell is assigned a discrete state
that is zero at the start and  evolves in discrete time steps by checking the local
neighborhood. More precisely, the state of a cell is incremented whenever it has a
neighbor with the same state situated upstream. The final state of a cell is thus equal to its position
along a chain of neighboring cells. Track candidates (TCs) are collected by starting from
the cells with the highest states and following the chain towards the origin of the track.
The resulting TCs may still share hits. For each of them, a quality indicator (QI) is computed
by a fast circle fit in the bending plane.  The final best set of non-overlapping TCs is
found by a Hopfield network that uses the QIs; this is the third filter stage.  

It is possible to iterate the track finder by using several sector maps corresponding to several
momentum ranges. In each iteration only TCs in a certain momentum range are found; the hits of
high-quality TCs are removed from the pool of available hits and are not used in the
following iteration. This feature was, however, not enabled in the test beam experiment. 

To determine the performance of the CA-TF, the data were scanned for events where the
following condition was met:
\begin{itemize}
 \item[$\hookrightarrow$] There was at least one hit loosely correlating with neighboring layers for the
three upstream telescope layers, the PXD and the SVD. The three downstream telescope layers had to provide
an additional correlation.
\end{itemize}
The surviving events were used for finding track candidates by
creating all possible hit combinations and fitting the result.

Only the best fit of an event was stored and used as the referenceTC
to measure the performance of the CA-TF. Since the combinatorial
problem of combining at least 8+3 layers is very severe, the fit
procedure was aborted if it took too long. Therefore not all events
fulfilling the conditions mentioned above provided reference TCs. As a
consequence, events with many hits did not give a reference TC,
although the CA-TF did. A straightforward definition of efficiency --
like the sum of the diagonal elements divided by the sum of all
elements -- is misleading, therefore a closer look is necessary. It
should be kept in mind that for the purposes of the HLT, the CA-TF
uses only hits in the SVD to find track candidates.

For a given run, a contingency table as shown in
Table~\ref{tab:contingency} can be created.  The sum
$n_{\mathrm{tot}}$ of all elements in the table is equal to the number
of events in the run.  The table can be used to define a measure of
efficiency of the CA-TF.
\begin{table}\caption{Contingency table.}\label{tab:contingency}
\centering
\begin{tabular}{|c|c|c|}
    \hline
      & no reference TC & reference TC \\
    \hline
    no TC from CA-TF  & $n_{00}$ & $n_{10}$ \\
    \hline
    TC from CA-TF   & $n_{01}$ & $n_{11}$ \\
    \hline
\end{tabular}
\end{table}
To deal with the special situation of the beam test two different types of efficiency were
used to benchmark the TF performance:
\begin{enumerate}
\item the number of events where both a reference TC and a CA-TC were
  present, divided by the number of events with a reference TC:
  $$\epsilon_1=\frac{ n_{11}} {n_{10} + n_{11}};$$
\item the number of events where the CA-TF found a TC, divided by the total number of
  events where a TC was found by either approach:
  $$\epsilon_2=\frac{n_{01} + n_{11}} {n_{\mathrm{tot}} - n_{00}}.$$
\end{enumerate}
The first definition $\epsilon_1$ ignores events in which no reference TC was found,
therefore high occupancy events are underrepresented. The second definition $\epsilon_2$ ignores
cases when the reference failed to find TCs and uses the reference
only to find events which can be excluded from the total number of events.
So neither efficiency can fully describe the situation by itself, and both
should be considered.

\begin{figure}[!t]
  \centering
  \includegraphics[width=\linewidth]{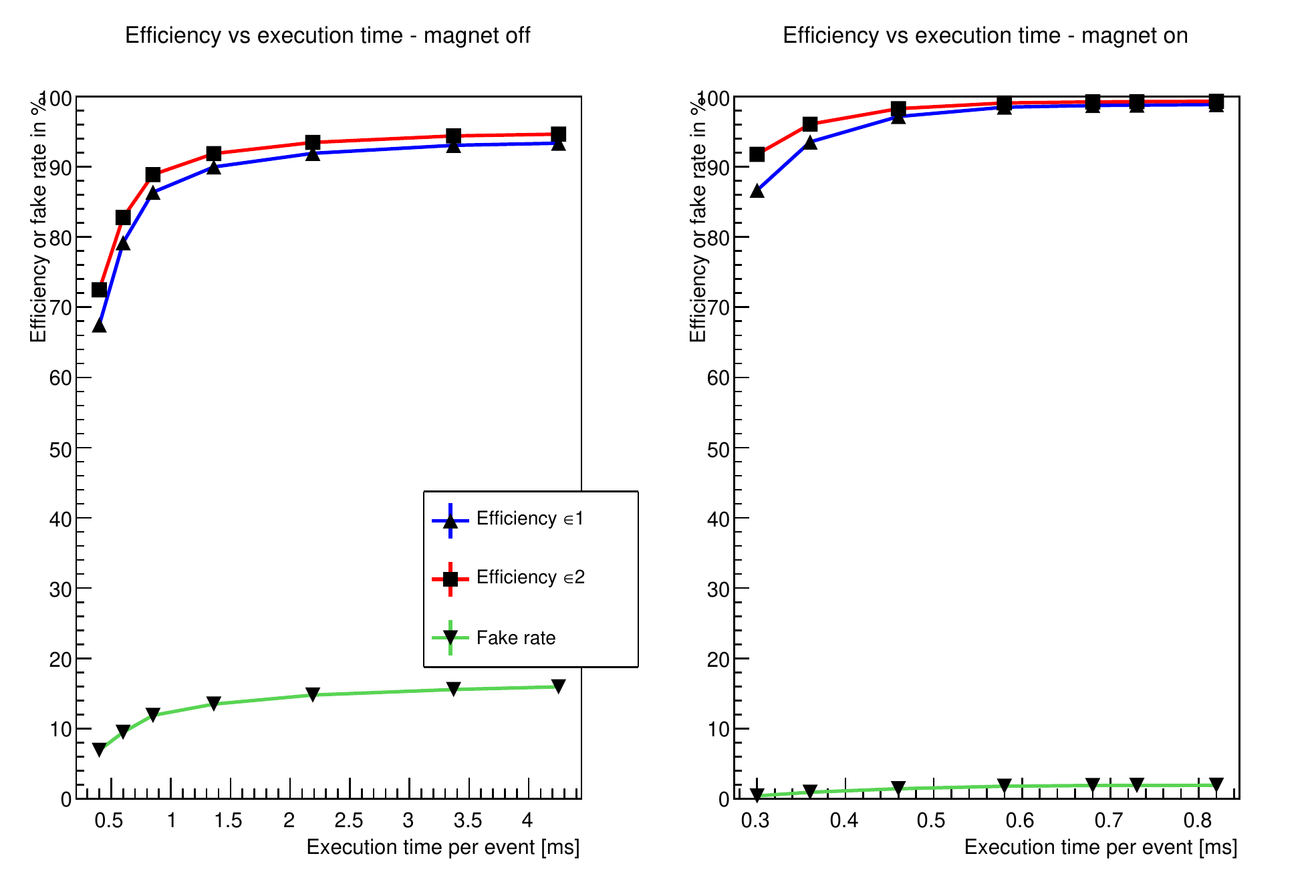}
  \caption{Left: run470, right: run510. Runs were analyzed using different thresholds (low: fast, high: slow) for the number of accepted hit combinations (cells).}
  \label{fig:speedEfficiency}
\end{figure}

Another relevant aspect is the fake rate. If we assume that there is only one real track in the event,
the fake rate (in percent) can be defined as 
$$ \frac{100}{n_{01} + n_{11}} \cdot (n_{\mathrm{CA\mbox{-}TC}} - (n_{01} + n_{11}))$$
where $n_{\mathrm{CA\mbox{-}TC}}$ ist the total number of TCs found by
the CA-TF.  This definition is conservative, as some of the additional
TCs are probably genuine secondary particles.  This is confirmed by
the fact that the fake rate is much smaller with the magnetic field
on, because most of the secondary particles created by passing through
the magnet coil are deflected by the field and do not reach the SVD
(see Fig.~\ref{fig:speedEfficiency}).

The TF was tested in two different settings; first, a run without magnetic field,
which allows checks for high occupancy for the reason mentioned above; second, a run with
a field of $1 \mathrm{T}$ to test the behavior under more realistic conditions.
Table~\ref{tab:runs} gives an overview over the settings used for the chosen runs.

\begin{table}\caption{Settings of the two runs used in the track finding studies.}\label{tab:runs}
\centering
\begin{tabular}{|c|c|c|c|}
    \hline
    run  & field strength & beam energy & events \\
    \hline
    470  & $0 \mathrm{T}$ & $3 \mathrm{GeV}$ & 4687 \\
    510  & $1 \mathrm{T}$ & $3 \mathrm{GeV}$ & 10282 \\
    \hline
\end{tabular}
\end{table}

\begin{figure}[!t]
  \centering
  \includegraphics[width=\linewidth]{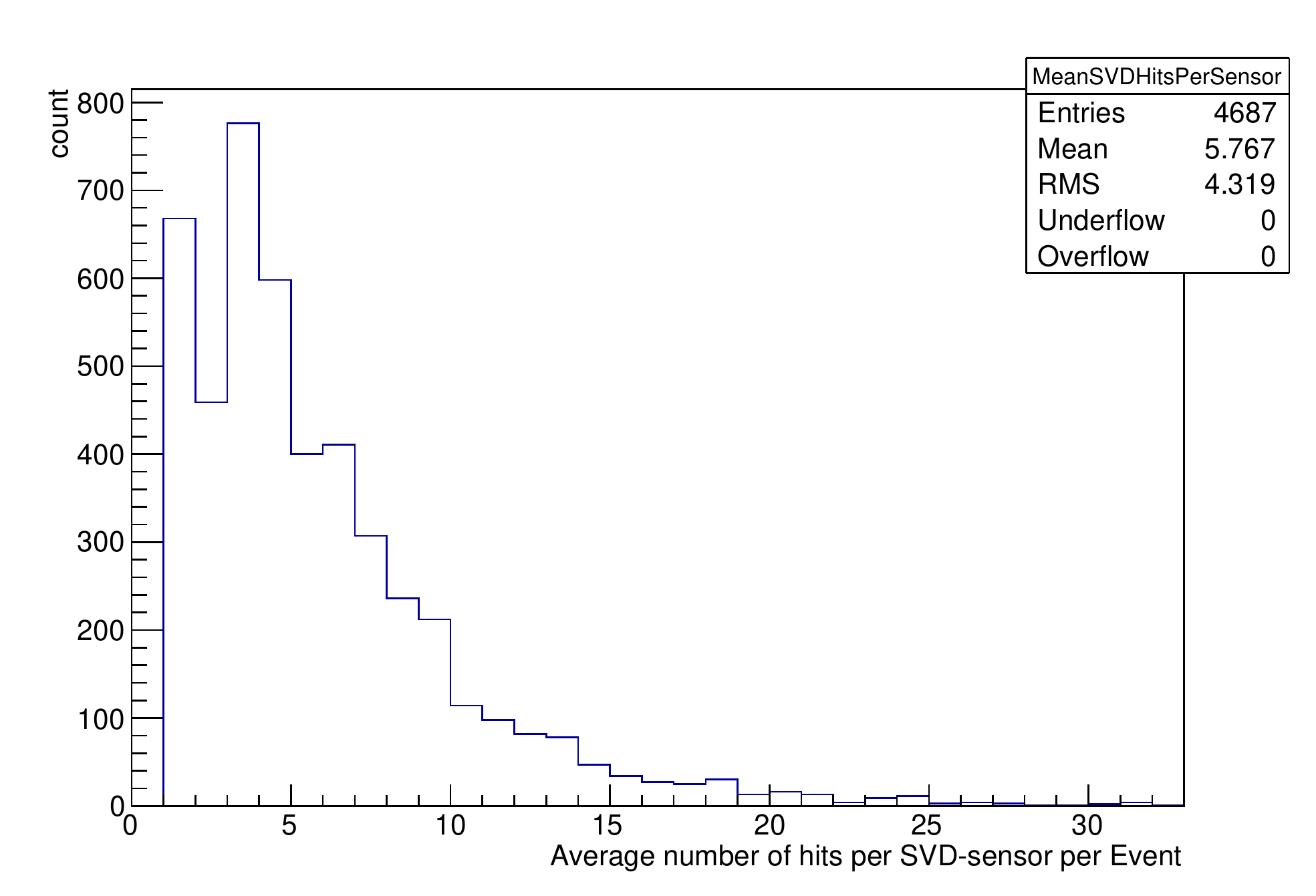}
  \caption{Average number of  hits per layer, run 470 (no magnetic field). }
  \label{fig:clusterMultiplicity470}
\end{figure}

\begin{figure}[!t]
  \centering
  \includegraphics[width=\linewidth]{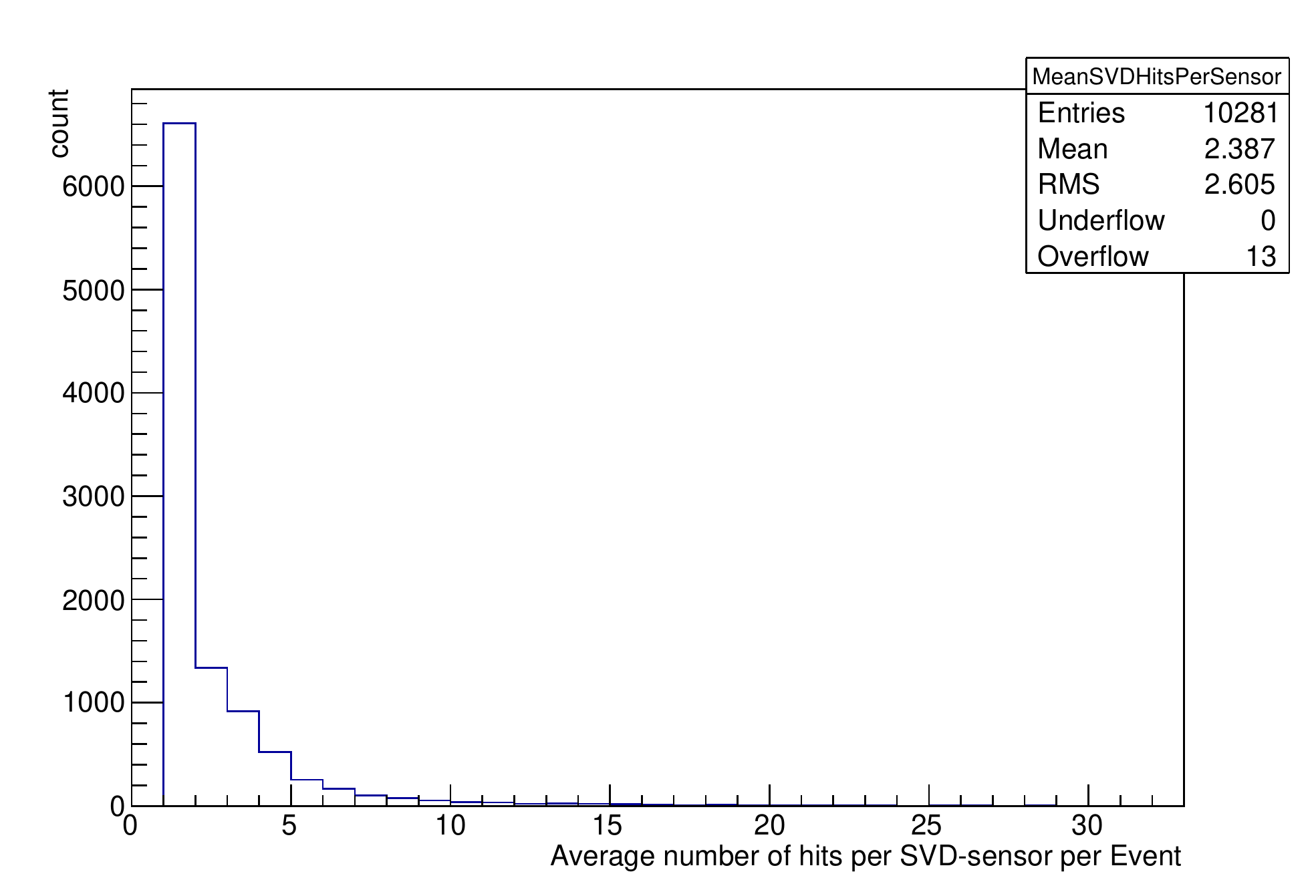}
  \caption{Average number of hits per layer, run 510 ($1\,\mathrm{T}$
    magnetic field). }
  \label{fig:clusterMultiplicity510}
\end{figure}

The frequency distributions of the average number of hits per layer in
the SVD are shown in Figs.~\ref{fig:clusterMultiplicity470}
and~\ref{fig:clusterMultiplicity510} for runs 470 and 510,
respectively. Taking the average over the layers as a rough benchmark,
run 470 with magnet off provided an average number of about 1100
possible hit combinations per event. Many secondary particles produce
a high rate of ghost hits. This background is a good test for high
track multiplicity.  Run 510 with magnet on provided an average number
of only 32 possible combinations per event. This has a major impact on
the execution time per event. During the beam test a maximum trigger
rate of about 1\,kHz limited the total time budget at the HLT
Demonstrator to about 40\,ms. To prevent bottle-necks for other
processing steps such as track fitting and ROI finding, the TF should
use only 10\% of the given budget, or $4\,\textrm{ms}$.
The speed of the CA-TF is mainly limited by the total number of
\textit{accepted} hit combinations per event, i.e., the ones that are
used as cells in the CA. To test the behavior of the CA-TF the chosen
runs were analyzed with the following thresholds on the number of
cells: 1250, 1000, 750, 500, 250, 125, 65. Events exceeding the
threshold are skipped. In Fig.~\ref{fig:speedEfficiency}, the
performance of the CA-TF is plotted as function of execution time per
event for both runs.  Using the highest threshold results in the best
efficiencies but requires the longest execution time per event. While
run 510 can be analyzed with the highest threshold within the timing
constraints, for run 470 the limit should be set to about 1000 in
order to stay below the limit of 4\,ms. In both runs the CA-TF found
TCs much more often than the reference, which can be partially
explained by the terminating condition for high occupancy cases with
the reference procedure. In the run with activated magnetic field the
effect of high occupancy is less dominating. Much fewer
reference TCs were generated because the downstream telescope sensors
(not required by the CA-TF) were not hit by particles that lost a
significant fraction of their momentum before entering the detector
volume (see Fig.~\ref{fig:mom-spectra}).

\subsection{Track Fitting}

\begin{figure}[!t]
  \centering
  \includegraphics[width=\linewidth]{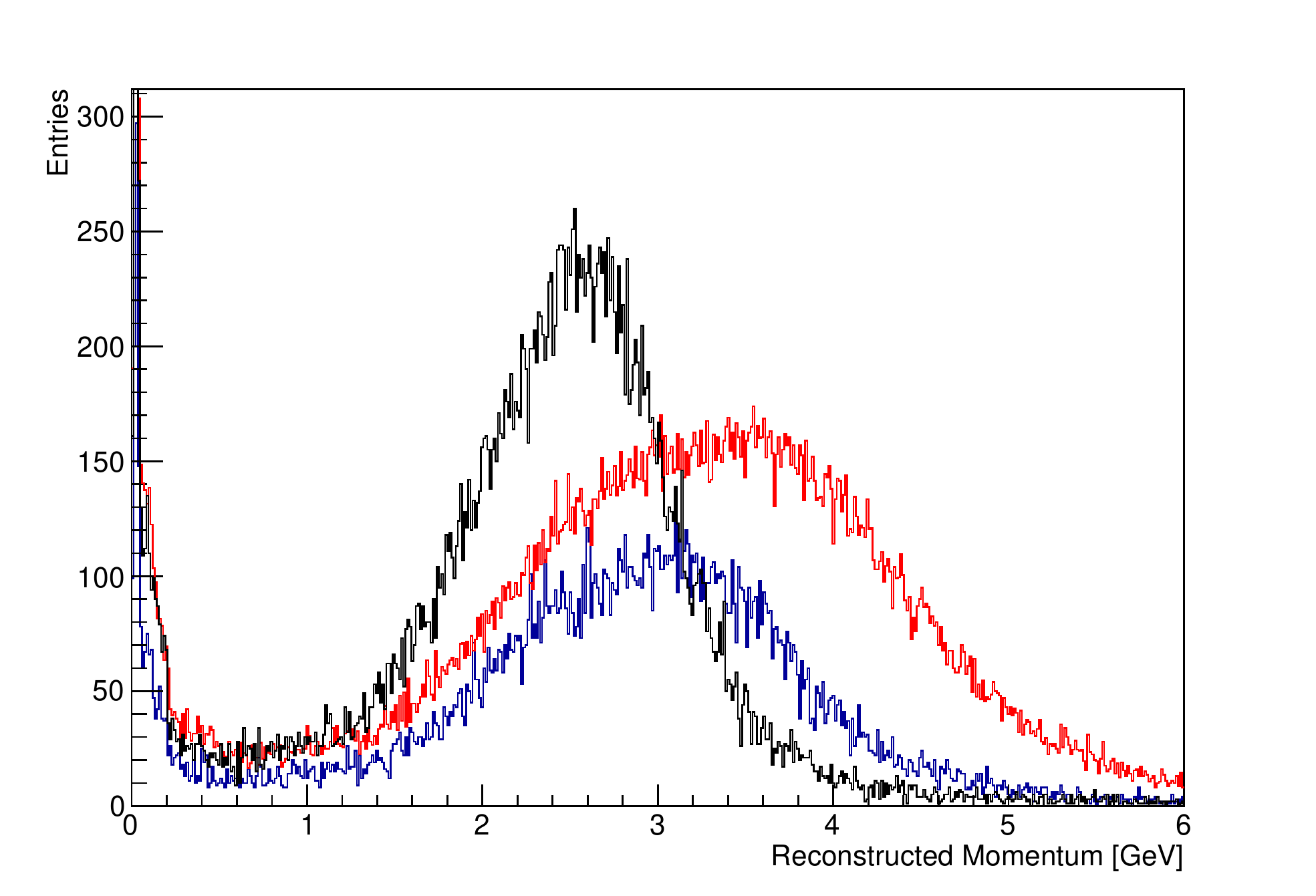}
  \caption{Measured momenta in GeV with three different beam settings.
    The curves are scaled to improve visibility.  The
    momentum spread is due to bremsstrahlung energy losses of the
    electrons while passing through the solenoid magnet.  Low-momentum
    tracks are bent away from the detector planes.}
  \label{fig:mom-spectra}
\end{figure}

The hit combinations together with an initial momentum estimate are
stored as track candidates.  These are then processed with a version
of the \genfit{} track fitting software~\cite{Hoppner:2009af}.  This
track fitting package has been largely overhauled to meet the
requirements of the \belle{} experiment.
\genfit{} implements a flexible framework for the modelling,
extrapolating and fitting of charged-particle trajectories in complex
detector environments.  For the purposes of definition of ROIs, the
track candidates are fitted with the standard Kalman fitter
algorithm~\cite{Fruhwirth:1987fm}.  For the purposes of detector
alignment the Generalized Broken Lines (GBL)~\cite{Kleinwort:2012np}
algorithm is employed offline.  Another track fitting algorithm
implemented in \genfit{} is the Deterministic Annealing Filter.  This
is the standard algorithm used in \belle{}.  The average time for the
fit of a single track in the current experimental setup was $<
1\,\textrm{ms}$ on a typical laptop.  Computing time requirements of
the HLT were thus easily fulfilled.  In Fig.~\ref{fig:mom-spectra}
reconstructed momentum spectra for runs with three different beam
momentum settings are shown.

\subsection{Determination of Regions of Interest}

Once tracks are fitted based on hits in the SVD planes, the definition
of the regions of interest (ROIs) takes place.  These are defined by
areas on the PXD planes which are likely to have been crossed by the
fitted tracks.  Once the ROIs are determined by the HLT, they are
forwarded to the Online Selector Node system (ONSEN) which holds the
PXD data until requested, and the ONSEN system in turn forwards the
PXD data contained within the ROIs to the second event builder (EVB2).
Here the pixel data is merged with the SVD data (and the data from the
other detectors in the full experiment) in order to assemble the
complete events which are then stored to disk.

The ROIs are found by the following procedure:
\begin{enumerate}
\item Find intercepts of all fitted tracks with the PXD planes.
\item Define rectangular arrays of pixels surrounding the intercept.
  The size of the array is determined by both the extrapolation errors
  and an allowance for systematic errors such as misalignment.
\end{enumerate}

\begin{figure}[!t]
  \centering
  \includegraphics[width=\linewidth]{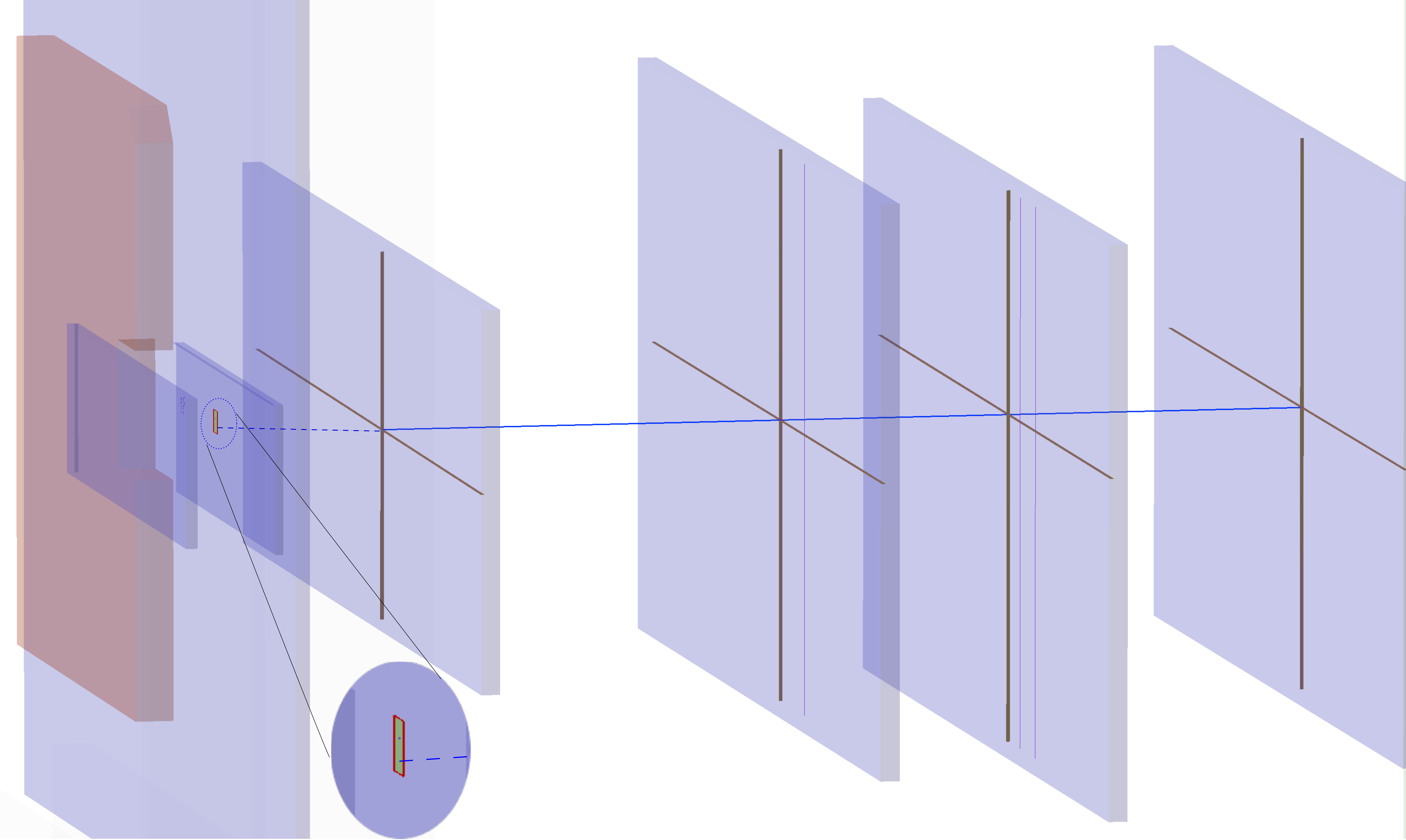}
  \caption{Illustration of the ROI definition process.  A track (blue,
    enters from the left) is fitted through the hits on the four SVD
    planes (four rightmost light blue planes).  Based on the backwards
    extrapolation (dotted blue) to the PXD plane, the ROI is defined
    (area with red boundary).  Indeed a hit is found within this area.
    The area surrounding the ROI is shown enlarged in the inset.}
  \label{fig:ROI}
\end{figure}

\begin{figure}[!t]
  \centering
  \includegraphics[width=\linewidth]{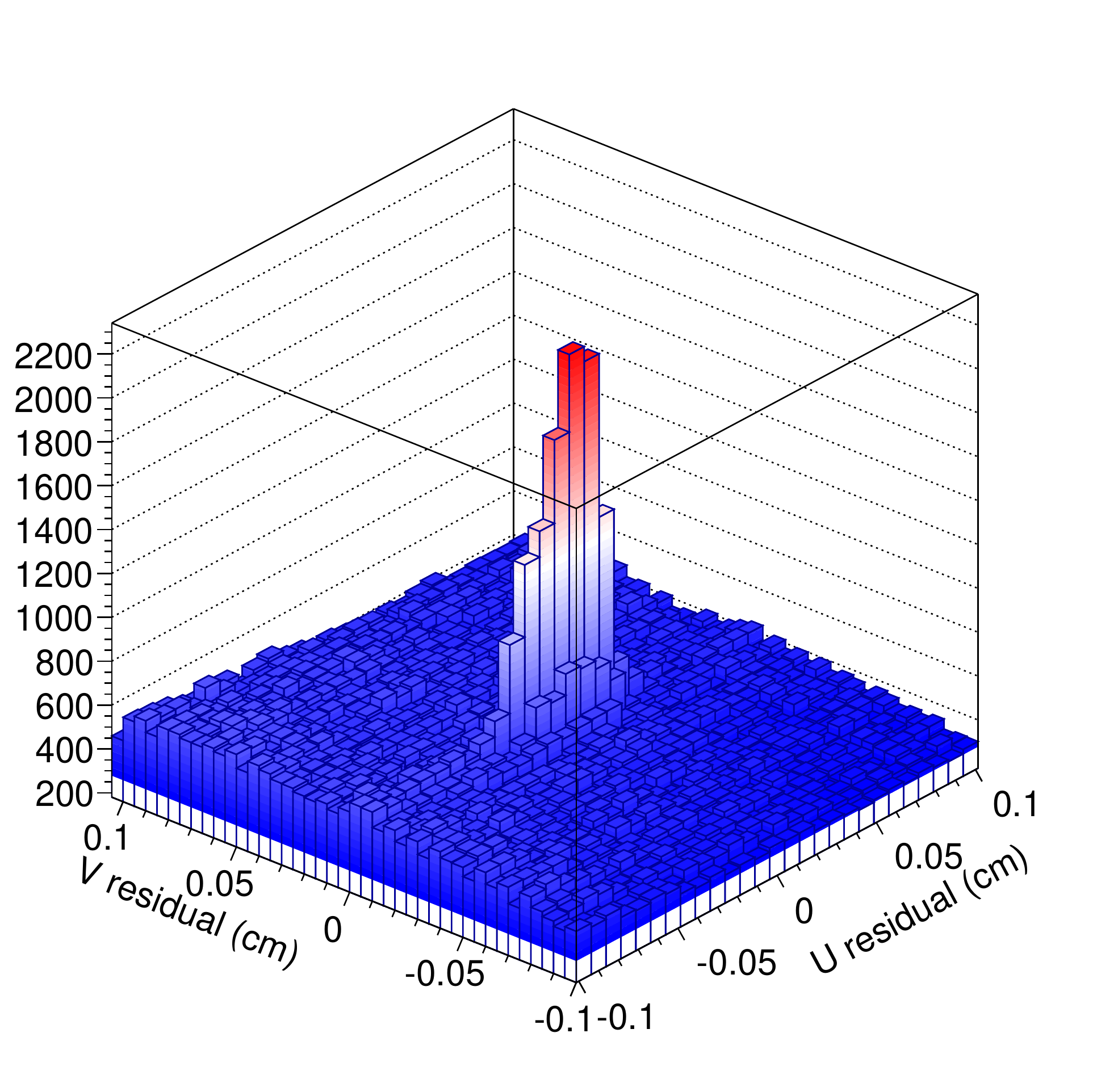}
  \caption{Residual between the intercept of the reconstructed
    SVD-track on the PXD plane and the actual PXD hits for
    $1\,000\,000$ test beam events.  The bins match the PXD pixels.  A
    Gaussian fit estimates the width of the central peak as $(32.3\pm
    0.4)\,\mu\textrm{m}$ and $(141\pm 2)\,\mu\textrm{m}$ in the $V$ and $U$
    directions, respectively.  Here $U$ denotes the bending direction.}
  \label{fig:2D-resid-lego}
\end{figure}

A high efficiency of the ROI determination is observed, leading to a
more than twenty-fold reduction of the amount of PXD data.
Communication of the ROIs to the ONSEN system could be established,
both transferring artificial ROIs (full detector plane, patterns) and
real ROIs determined in the above manner.  The procedure is
illustrated for an example event in Fig.~\ref{fig:ROI}.  A
distribution of two-dimensional offsets between the measured pixel
hits and the calculated intercepts is shown in
Fig.~\ref{fig:2D-resid-lego}.


\subsection{Alignment}

For the purpose of alignment of the detector setup, the default
alignment procedure developed for the \belle{} vertex detector was
applied.  The procedure is fully integrated into the \belle{} software
and utilizes the Millepede II algorithm described in
Ref.~\cite{Blobel:2006yh}.  Data for Millepede~II are prepared by a
refit of reference tracks (provided by the track finder) using the
implementation of the General Broken Lines (GBL)
algorithm~\cite{Kleinwort:2012np} in the GENFIT package.  In GBL,
multiple scattering is taken into account representing thick
scatterers by two equivalent thin scatterers at detector planes and in
between.

\begin{figure}[htbp!]
  \centering

  \includegraphics[width=\linewidth]{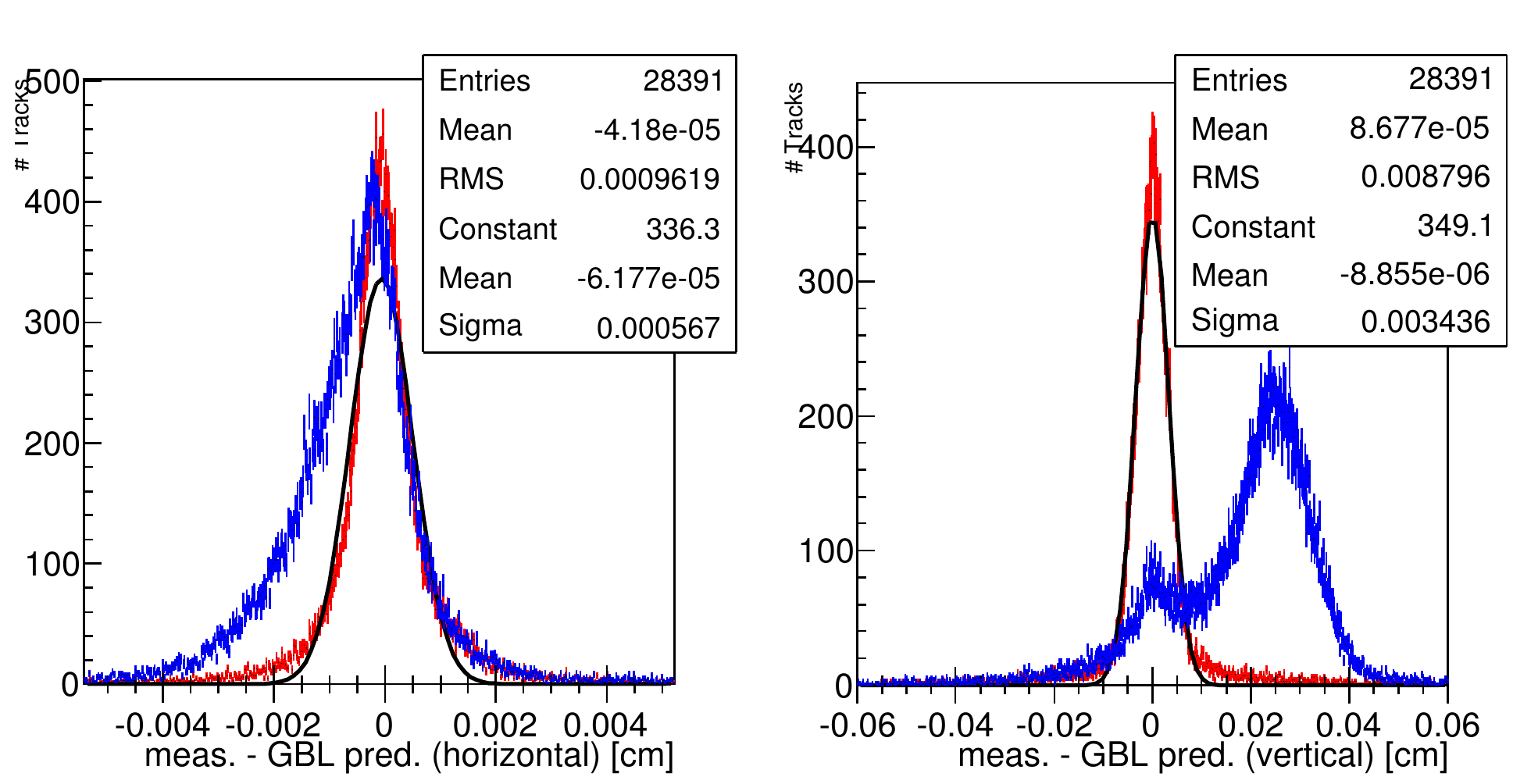}
  \caption{GBL refit residuals in $V$ (left) and $U$ (right)
    directions of the second SVD sensor using nominal (blue) and
    aligned (red) geometry. Parameters of a Gaussian fit (black curve)
    to the red histogram are given.}
\label{fig:alignment}
\end{figure}

The alignment procedure is successfully applied for data obtained with
and without magnetic field.  Alignment corrections determined for
sensor displacements and rotations in their planes have typical
uncertainties below $3\,\mu{}\textrm{}m$ / $0.5\,\textrm{mrad}$ and
differ usually by less than $1\,\textrm{mm}$ / $10\,\textrm{mrad}$
from the nominal geometry. The only exception is the PXD sensor
shifted by about 5mm.  As an example, in Fig.~\ref{fig:alignment}
changes of residual distributions of the second SVD sensor after
alignment in the magnetic field is presented.  Clear improvements of
residuals after the alignment procedure are observed.

\section{Summary}

During the test beam campaign which concluded in January 2014 the
readout and data reduction scheme of the \belle{} vertex detector
could be proven.  A number of milestones were achieved:
\begin{enumerate}
\item simultaneous operation of SVD and PXD detectors;
\item common readout of all subsystems with the final \belle{} architecture;
\item online, real-time processing of SVD data with the full
  reconstruction software also used for offline;
\item usage of the \belle{} software framework's network data
  distribution and parallel processing capabilities;
\item cellular automaton based track finding;
\item readout of the PXD could be successfully be driven by the SVD
  data;
\item processing steps such as alignment could successfully be
  performed.
\end{enumerate}

\section*{Acknowledgment}
This work would not have been possible without hard work by the Belle
II PXD, SVD and DAQ test beam teams, led by C.~Marinas, C.~Irmler and
R.~Itoh, respectively, nor would it have been possible without the
DESY test beam facilities.


\end{document}